\documentstyle[epsfig,12pt]{article}

\newcommand{\be}{\begin{equation}}
\newcommand{\ee}{\end{equation}}
\newcommand{\ba}{\begin{eqnarray}}
\newcommand{\ea}{\end{eqnarray}}
\newcommand{\bd}{\begin{displaymath}}
\newcommand{\ed}{\end{displaymath}}

\def\thalf{{\textstyle{\frac{1}{2}}}}

\def\oneth{{\textstyle{\frac{1}{3}}}}
\def\twoth{{\textstyle{\frac{2}{3}}}}
\def\oneqt{{\textstyle{\frac{1}{4}}}}
\def\ttqt{{\textstyle{\frac{3}{4}}}}

\title{
{\bf Formation of Gapless Phases of $K^0$ Condensed Color-Flavor Locked 
Superconducting Quark Matter}}
\author{{ Xiao-Bing Zhang$^{1,2}$ and J. I. Kapusta$^{2}$} \vspace*{0.1in}\\
$^1${\it Department of Physics, Nankai University}\\
 {\it Tianjin 300071, China}\\
$^2${\it School of Physics and Astronomy, University of Minnesota}\\
 {\it Minneapolis, Minnesota 55455, USA}}

\date{}

\begin{document}

\maketitle

\begin{abstract}
Electric and color neutral solutions, and the critical conditions for the 
formation of gapless color superconductors, are investigated in $K^0$ condensed 
color-flavor locked quark matter for nonzero strange quark mass. We show that as 
the strange quark mass increases, gapless modes for up-strange quark pairing 
occur first, followed by down-strange quark pairing.  The behavior of the gaps, 
the dispersion relations, and the thermodynamic potential are all found as 
functions of the strange quark mass on the basis of a Nambu Jona-Lasinio type 
model.  To a high degree of accuracy, they are presented as relatively simple 
elementary functions.  This allows for easy computation for any reasonable range 
of baryon chemical potential and strange quark mass.
\end{abstract}

\vspace{0.5cm} {PACS numbers: 25.75.Nq, 12.39.Fe, 12.38.-t}

\newpage

\section{Introduction}

Dense quark matter has been an exciting topic for theoretical study ever since 
it was realized that quark color superconducting phases could form the lowest 
energy states and could be reliably computed at very high density within QCD. 
For three flavors, the original color and flavor $SU(3)_{\rm color}\times 
SU(3)_L \times SU(3)_R$ symmetries of QCD are broken down to the diagonal 
subgroup $SU(3)_{{\rm color}+L+R}$ at very high baryon density \cite{alf}. Quark 
matter with such a particular symmetry pattern is called color-flavor locked 
(CFL) matter and is widely believed to be the densest phase of strongly 
interacting matter \cite{alf03}. In the situation where the strange quark mass 
$m_s$ is large and/or the quark chemical potential $\mu$ (which is 1/3 of the 
baryon chemical potential) is small, the so-called gapless color-flavor locked 
phase (gCFL) has been predicted \cite{liu,alf04,alf05}. Similar to the gapless 
phase of two-flavor color superconductor \cite{huang}, gCFL is triggered by 
mismatches between the chemical potentials for paired quasi-quarks. There the 
mismatch in the flavor (say $d$ and $s$) chemical potentials is characterized 
directly by the quantity $m_s^2/2\mu$ while that in the color (say $b$ and $g$) 
chemical potentials is related to $m_s^2/2\mu$ via the electric/color neutrality 
in CFL matter \cite{alf04}. As long as $m_s^2/2\mu$ is large enough, there exist 
unpaired (gapless) modes for $bd-gs$ pairing, and the resulting gCFL phase 
becomes energetically favorable. Recently the Meissner masses for some gluons in 
this phase were found to be imaginary so that gCFL is actually unstable in the 
chromomagnetic sense \cite{cas05,fuku,alfjp05}. The presence of a chromomagnetic 
instability is a serious problem.  Hence the gapless phase of CFL matter needs 
to be studied further.

There is another kind of less-symmetric phase of CFL matter that was predicted 
when $m_s \neq 0$. At the leading order, it is convenient to regard the effect 
of $m_s^2/2\mu$ as an effective chemical potential associated with strangeness, 
namely
\be 
\mu_S = \frac{m_s^2}{2\mu} \, . \label{muk}
\ee
As one of the pseudo-Goldstone bosons related to excitations of the CFL 
superconducting quark matter (not to be confused with excitations of the QCD 
vacuum with the same quantum numbers), the neutral kaon mode $K^0$ has a 
chemical potential $\mu_{K^0}=\mu_S$. The mass of this mode, ignoring instanton 
effects, is \cite{sch}
\be
m_K^0 = \frac{\Delta}{\pi f_{\pi}} \sqrt{3m_u (m_s + m_d)}
\ee
where $\Delta$ is the superconducting gap, $m_u$, $m_d$ and $m_s$ are the quark 
masses, and $f_{\pi}$ is the pion-decay constant in the CFL matter.  The latter 
is given by \cite{fpi}
\be
f^2_{\pi} = \frac{21 - 8 \ln 2}{36 \pi^2} \mu^2 \approx (0.2086 \mu)^2\, .
\ee
It measures the strength of the coupling of the pion (in CFL matter) to the 
axial vector current.  When $\mu_S$ reaches the mass of the neutral kaon-mode, 
the $K^0$ condenses in CFL matter \cite{sch,kr}. This phase is referred to as 
CFL$K^0$.  Assuming $m_d \ll m_s$, the condition that no $K^0$ condensation 
takes place is
\be
m_s < 3.034 \left( m_u \Delta^2 \right)^{1/3}
= 44.4 \left[ \left( \frac{m_u}{5 \,{\rm MeV}} \right) 
\left( \frac{\Delta}{25 \,{\rm MeV}} \right)^2 \right]^{1/3} \,\, {\rm MeV} \, .
\ee
Taking typical current-quark values of the masses, namely $m_u = 5$ MeV and $m_s 
= 110$ MeV, kaon condensation will occur for reasonable values of the gap.  In 
this paper we will generally ignore the up and down quark masses in comparison 
with the strange quark mass.

Starting with the ideal situation of $m_s=0$ and raising $m_s^2/2\mu$ gradually, 
one expects the CFL matter to be disrupted, at first by the presence of $K^0$ 
condensation, and then by the appearance of gapless modes. In the CFL matter 
with $K^0$ condensation, the formation of gapless modes has been reexamined by 
using an effective theory in Ref. \cite{kry1,kry2} and by using the NJL model in 
Refs. \cite{forbes,buba}. It was found that gapless pairing is delayed with 
respect to that in the conventional CFL matter. More recently, it has been 
suggested that the chromomagnetic instability in the gapless phase can be 
resolved by the formation of the so-called kaon supercurrent state \cite{sch06}.

Motivated by these results, we investigate the formation of gapless modes in the 
CFL$K^0$ environment. First, we consider the electric/color neutral solution of 
CFL$K^0$ in the presence of an electron chemical potential and point out that 
the above-mentioned delay is a direct consequence of the deviation of the 
CFL$K^0$ neutral solution from the CFL one. Second, it is found that gapless 
modes for $bu-rs$ pairing occur first, followed by $bd-gs$ in the $K^0$ 
condensed environment. Based on this feature, the resulting gapless phase 
(termed as gCFL$K^0$ below) is studied, including its electric/color neutrality 
and gap variation.  The solutions can be found analytically to very good 
approximation, obviating the need for a complicated numerical minimization of 
the thermodynamic potential with respect to three chemical potentials plus three 
gaps.  Finally, the stability of the gCFL$K^0$ phase is examined qualitatively; 
it is argued that the gCFL instability might be removed either partially or 
even totally.

\section{Electric/Color Neutrality in Superconducting States}

At asymptotically high baryon density the gap in CFL matter can be computed 
using weak coupling methods.  The result is \cite{QCDgaps}\be
\Delta = \frac{512 \pi^4}{2^{1/3}} \left(\frac{2}{N_f}\right)^{5/2}
\frac{\mu}{g^5}
\exp\left[ - \frac{(\pi^2+4)}{8} \right] 
\exp\left( - \frac{3\pi^2}{\sqrt{2} g} \right)
\ee
where $N_f$ is the number of flavors.  (In this paper we consider $N_f=3$ only.)  
In the asymptotic limit, $\mu \gg \Lambda_{\rm QCD}$, it is also true that $\mu 
\gg m_s$ or, equivalently, $\mu_S \ll \Delta$.  The only location in the known 
universe where superconducting quark matter might appear is in neutron stars 
where the baryon chemical potential is not asymptotically large.  Then $\mu_S$ 
is not negligible, and furthermore weak coupling methods in the gauge coupling 
$g$ are not quantitatively reliable.  Therefore different groups have used other 
methods to understand what happens in the non-asymptotic regime.  These include:  
(i) variations on the Nambu Jona-Lasinio model, which allow a determination of 
the gaps as functions of density at the expense of one or more cut-off 
parameters, (ii) effective Lagrangians employing the relevant collective degrees 
of freedom and with coefficients matched to QCD, but which assume a given value 
for the gap, and (iii) so-called model independent approaches that also assume 
that the gap or gaps are determined by other means and that count states, 
degrees of freedom, and balance electric and color charges.  These different 
approaches all tend to agree when they have a region of overlap.

Stable bulk matter must be electrically neutral and must have zero net color.  
Perfect CFL matter with zero quark masses satisfies these requirements without 
the need for electrons.  Alford and Rajagopal \cite{alfj02} addressed this issue 
in the context of a nonzero strange quark mass by using an essentially
model-independent approach.  They introduced chemical potentials $\mu_e$, 
$\mu_3$ and $\mu_8$ for the electrons (or negative of the electric charge $Q$) 
and for the color generators $T_3$ and $2 T_8/\sqrt{3}$, respectively.  (The 
factor of $2/\sqrt{3}$ arises because they used a non-standard form of the 
generator $T_8$ which was more convenient for their purposes.)  Different 
flavors and colors of quarks then have different chemical potentials; see Table 
I.  The thermodynamic potential $\Omega$ must satisfy
\be
\frac{\partial \Omega}{\partial \mu_e}=\frac{\partial \Omega}{\partial \mu_3}
=\frac{\partial \Omega}{\partial \mu_8}=0 \, .
\label{neu}
\ee
They found the solution
\be 
\mu_3=\mu_e \, , \,\,\,\,\,\,
\mu_8=\thalf \mu_e-\mu_S  \label{ncfl}
\ee
to first order in $\mu_s/\mu$.  (The electron chemical potential is not 
determined by this calculation.)  This solution is equivalent to requiring that 
the quarks that should pair have the same ``Fermi momentum" to the extent that 
such a quantity has meaning in superconducting matter.  To order $m_s^4$ (and to 
zero order in $g^2$) the difference in thermodynamic potential between the CFL 
phase and ordinary unpaired quark matter is
\be
\Omega^{\rm neutral}_{\rm CFL} = \Omega^{\rm neutral}_{\rm unpaired}
+ \frac{3\mu^2}{4 \pi^2} \left( \mu_S^2 - 4 \Delta^2 \right) \, .
\label{deltaCFL}
\ee
The CFL phase is preferred over the unpaired phase if the gap is large enough, 
namely if $\thalf \mu_S < \Delta$.  Another important feature of CFL is that its 
thermodynamic potential is independent of
\be 
\mu_{\tilde{Q}}=-{\textstyle\frac{4}{9}}
\left(\mu_e+\mu_3+\thalf \mu_8\right) \, ,
\label{Q}
\ee
since none of the CFL pairings break the rotation
${\tilde{Q}}=Q-T_3-\frac{1}{\sqrt{3}}T_8$. In this sense, CFL matter is not 
merely an electric insulator \cite{rw01} but also a ${\tilde{Q}}$ insulator 
\cite{alfj02}.  Enforcing neutrality at higher order requires that no electrons 
be present so that $\mu_e = 0$.

As mentioned in the introduction, kaon condensation will occur when the kaon 
chemical potential equals its mass.  (We must keep in mind that familiar names 
like kaon refer to collective excitations of the color superconducting state and 
not to excitations of the QCD vacuum.)  Since we are taking the up and down 
quark masses to be zero (actually only $m_u=0$ is required in this regard) kaon 
condensation will always occur, and CFL$K^0$ matter should have a lower free 
energy than CFL.  In the presence of $K^0$ condensation, the ground state of 
Nambu-Goldstone bosons takes non-trivial values.  By treating the gauge fields 
as dynamical degrees of freedom, Kryjevski has shown that there exist nonzero 
expectation values for the gluon fields in CFL$K^0$ matter \cite{kry03}. He 
found that electric and color neutrality leads to constant time components of 
two color-diagonal gluon fields, which turn out to be
\be 
gA^0_3=\frac{m_s^2}{4\mu} \, ,\,\,\,\,\,\,\, gA^0_8
=\frac{m_s^2}{4\sqrt{3}\mu} \, .
\label{a38}
\ee
Constant values of the time-components of gauge fields play the mathematical 
role of shifts in the chemical potentials.  In this case it corresponds to
\be
\mu_3= - \thalf\mu_S \, , \,\,\,\,\,\,
\mu_8= - \oneqt\mu_S  
\ee
when the condition $\mu_e=0$ is used.  (Neutral kaon condensation does not 
require the presence of electrons for electric neutrality.)  These results were 
confirmed by Kryjevski and Yamada \cite{kry2}, as were the results of Alford and 
Rajagopal.  A comparison of the effective potentials for the quarks in the CFK 
and CFL$K^0$ states are given in Table I.  The thermodynamic potential in the 
CFL$K^0$ state is lower than that of the CFL state (including the strange quark 
mass effects, electric and color neutrality) by an amount $\thalf f_{\pi}^2 
\mu_S^2$ \cite{sch,kry2}.  Together with (\ref{deltaCFL}) this means that
\be
\Omega^{\rm neutral}_{{\rm CFL}K^0} = \Omega^{\rm neutral}_{\rm unpaired}
+ \frac{3\mu^2}{4 \pi^2} \left( \mu_S^2 - 4 \Delta^2 \right) 
- \frac{1}{2} f_{\pi}^2 \mu_S^2\, .
\label{deltaCFLK}
\ee
The condition for pairing is now somewhat relaxed to $0.4224 \mu_S < \Delta$ as 
opposed to $0.5 \mu_S < \Delta$.

\section{Dispersion Relations}

The modes of excitation in superconducting quark matter are sometimes labeled by 
the quark degrees of freedom and sometimes by the baryon degrees of freedom.  
The correspondence is related to the pairing ansatz
\be
{\langle q^\alpha_i C\gamma_5
q^\beta_j\rangle} \sim \Delta_I \epsilon^{\alpha\beta
1}\epsilon_{ij1} + \Delta_{II} \epsilon^{\alpha\beta 2}\epsilon_{ij2} +
\Delta_{III} \epsilon^{\alpha\beta 3}\epsilon_{ij3} \, .
\ee
(Here $(i,j)$ and $(\alpha,\beta)$ denote the flavor indices
$(u,d,s)$ and the color indices $(r,g,b)$, respectively, and the gap parameters 
$\Delta_I$, $\Delta_{II}$ and $\Delta_{III}$ are approximately equal to 
$\Delta_0$ in 
the absence of gapless phenomena.) For example, a linear combination of $\langle 
ru-gd \rangle bu$ and $\langle gu-rd \rangle bu$, where the angular brackets 
denote a pairing, represent an object with baryon number 1, electric charge 1, 
and zero net color, in other words a proton.  Hence a proton is associated with 
$bu$ quarks.  Similarly the neutron, $\Sigma^+$, $\Sigma^-$, $\Xi^0$ and 
$\Xi^-$ are associated with the $bd$, $gu$, $rd$, $gs$ and $rs$ quarks,  
respectively.  A $\Lambda^0$, $\Lambda^8$, or $\Sigma^0$ would be represented by  
some linear combination of $ru$, $gd$ and $bs$ quarks.  These associations are 
given in Table IV. 

In the CFL phase with massless up, down and strange quarks the three gaps are 
equal: $\Delta_I = \Delta_{II} = \Delta_{III} = \Delta_0$.  The dispersion 
relations for the energy as a function of momentum take the following forms.  
There are 8 modes with energy
\be
E = \sqrt{(p-\mu)^2 + \Delta_0^2}
\ee
and 1 mode with energy
\be
E = \sqrt{(p-\mu)^2 + (2\Delta_0)^2} \, .
\ee
The mode with gap $2\Delta_0$ is a mixture of $ru$, $gd$ and $bs$ quarks.

In the CFL phase with strange quarks of nonzero mass the dispersion relations 
are more complicated because of the necessity of nonzero effective chemical 
potentials that ensure color neutrality.  These are found as the zeros of a 
determinant of a $9\times 9$ matrix \cite{alf04}.  There are a pair of modes 
associated with each of the pairings $bd-gs$, $bu-rs$ and $gu-rd$ described by 
the gaps $\Delta_I$, $\Delta_{II}$ and $\Delta_{III}$, respectively.  We shall 
refer to 
these as the I, II and III channels for brevity.  In this phase all three gaps 
are equal to a common value denoted by $\Delta$, which in general is not equal 
to $\Delta_0$ when the strange quark mass is nonzero.  The pair of dispersion 
relations for channel I can be written as
\be
E = \sqrt{(p-\bar{\mu})^2 + \Delta^2} \pm \mu_S \, .
\ee
Here $\bar{\mu} \equiv \mu - \frac{1}{3} \mu_S$ is the average chemical 
potential of the pair of quarks involved and $\mu_S$ is one-half of the 
difference of their chemical potentials; see Table II.  The same is true in 
channel II.  The pair of dispersion relations for channel III are
\be
E = \sqrt{(p-\bar{\mu})^2 + \Delta^2} \, .\ee
The other three modes are linear combinations of $ru$, $gd$ and $bs$ quarks.  
There are 2 modes with energy
\be
E = \sqrt{(p-\bar{\mu})^2 + \Delta^2}
\ee
and 1 mode with energy
\be
E = \sqrt{(p-{\bar\mu})^2 + (2\Delta)^2} \, .
\ee

In the CFL$K^0$ phase it is sometimes convenient to use the notation of baryons 
instead of quarks.  The translation between the two is given in Table IV.  The 
dispersion relations are different from the CFL matter not only because of the 
difference in effective chemical potentials but also because of flavor required 
by the symmetries of QCD \cite{kry1,kry2}.  There is a pair of modes in channel 
I with 
energy given by
\be
E = \sqrt{\left(p-\bar{\mu}\right)^2 + \Delta^2} 
\pm \frac{1}{2} \mu_S \, .
\ee
There is a pair of modes in channel II with energy given by
\be
E = \sqrt{\left[p-\left(\bar{\mu} - \oneqt \mu_S\right)\right]^2 + \Delta^2} 
\pm \frac{3}{4} \mu_S \, .
\ee
There is a pair of modes in channel III with energy given by
\be
E = \sqrt{\left[p-\left(\bar{\mu} + \oneqt \mu_S\right)\right]^2 + \Delta^2} 
\pm \frac{1}{4} \mu_S \, .
\ee
The neutral baryons $n$, $\Xi^0$, $\Sigma^0$, $\Lambda^0$ and $\Lambda^8$ (or 
$bd$, $gs$ and $ru$, $gd$, $bs$) mix among themselves to produce the following 
dispersion relations.  There is a pair of modes with energies
\be
E = \sqrt{(p-\bar{\mu})^2 + \Delta^2} \, ,
\ee
and 1 mode with energy
\be
E = \sqrt{(p-\bar{\mu})^2 + (2\Delta)^2} \, .
\ee
   
\section{Transition to a Gapless Phase}

As pointed out in Refs.\cite{alf04,huang}, the gapless phase arises when the 
energy of one of the collective modes becomes zero.  As one goes deeper into 
that phase there will be a finite window of momentum for which the energy of the 
mode would be negative.  This is the blocking region. One needs to find a 
solution to the combined set of electric/color neutrality conditions and gap 
equations.  When gapless modes appear, the gaps for various pairings separate 
from each other and their values should be solved from three gap equations, 
namely ${\partial \Omega}/{\partial\Delta_I} =
{\partial \Omega}/{\partial \Delta_{II}}= {\partial\Omega}/{\partial 
\Delta_{III}}=0$. 
If only one mode becomes gapless, it is natural to focus on variations of that 
gap from the original value $\Delta_0$ while ignoring variations in the other 
two gaps.  If two modes become gapless simultaneously then we should consider 
variations of two gaps while ignoring variations of the third.

The gapless phase will arise when the mismatch between the chemical potentials, 
$\delta \mu$, of two quarks exceeds the gap.  These mismatches are displayed in 
Table II for both the CFL and the CFL$K^0$ phases.  See also the previous 
section.  The first mode to go gapless in CFL occurs when $\mu_S = \Delta_0$ and 
is associated with both $\Delta_I$ and $\Delta_{II}$.  The first mode to go 
gapless 
in CFL$K^0$ occurs when $\mu_S = 4\Delta_0/3$ and is associated with 
$\Delta_{II}$.  
We will focus on gCFL$K^0$ since we already know that CFL$K^0$ is favored over 
CFL when $\mu_S > 0$.

In Nambu Jona-Lasinio type models the thermodynamic potential is written as
\be
\Omega = \Omega_0 + \frac{\Delta_I^2 + \Delta_{II}^2 + \Delta_{III}^2}{G}
- \sum_{j=1}^{18} \int \frac{d^3p}{(2\pi)^3} \left| E_j(p) \right|
- \frac{\mu_e^4}{12\pi^2} \, ,
\label{basicNJL}
\ee
where $\Omega_0$ is a constant adjusted to make the energy of the true vacuum 
zero.  The constant $G$ is a coupling constant.  The sum runs over 3 flavors, 3 
colors, and particle plus antiparticle; spin is already included.  The energies 
for the particles are generically written as
\be
E_j(p) = \sqrt{(p-\bar{\mu}_j)^2 + \Delta_j^2} + \delta \mu_j \, .
\ee
The chemical potentials $\bar{\mu}_j$ and $\delta \mu_j$ are linear combinations 
of $\mu$, $\mu_3$, $\mu_8$ and $\mu_e$.  Anti-particles have chemical potentials 
opposite to particles.  The momentum integrals are quartically divergent; 
usually they are regulated with a hard momentum cut-off $\Lambda$.  In some 
sense this approximates asymptotic freedom of QCD in the context of this model.  
The exact expression for the integral with such a cut-off is given in the 
Appendix for completeness.  The approximate expression which includes terms that 
go as positive powers of $\Lambda$ or which remain finite as $\Lambda 
\rightarrow \infty$ is also given in the appendix.  There are several terms that 
are odd in the chemical potentials.  These terms cancel when adding particles 
and anti-particles, so in this sense the inclusion of anti-particles simplifies 
the analysis.  Keeping only those terms which are finite as $\Lambda \rightarrow 
\infty$ gives us
\bd
\Omega = \Omega_0 - \frac{9}{4\pi^2} \Lambda^4 - \frac{\mu_e^4}{12\pi^2}
+ \frac{\Delta_I^2 + \Delta_{II}^2 + \Delta_{III}^2}{G}
\ed
\be 
- \frac{1}{4\pi^2} \sum_{j=1}^9 \left[ \Lambda^2 \Delta_j^2
+ \frac{1}{2} \Delta_j^2 \left( 4\bar{\mu}_j^2 - \Delta_j^2 \right)
\ln \left( \frac{2\Lambda}{\Delta_j} \right)
+ \frac{1}{3} \bar{\mu}_j^4 - 2 \bar{\mu}_j^2 \Delta_j^2 
+ \frac{1}{8} \Delta_j^4 \right] \, .
\label{LargeL}
\ee
The term proportional to $\Lambda^4$ is absorbed into the vacuum energy.  The 
terms quadratic and logarithmic in $\Lambda$ are usually accommodated as 
follows: One sets the strange quark mass to zero and solves for the common gap 
$\Delta_0$ at some reference chemical potential $\mu$ for some fixed values of 
$G$ and $\Lambda$.  A change in $\Lambda$ is accompanied by a corresponding 
change in $G$ such that the gap does not change.  Generally it is found that the 
results are not sensitive to the specific choice of $\Lambda$ so long as it is 
large compared to $\mu$.

\subsection{Ideal CFL}

Consider the ideal CFL matter with massless quarks.  There are 8 modes with gap 
$\Delta$ and one mode with gap $2\Delta$.  These gaps are predicted to be very 
small compared to the chemical potential so the terms in the thermodynamic 
potential of order $\Delta^4$ are normally dropped.  The $\Omega_0$ and the $G$ 
get renormalized as
\ba
\Omega_0 - \frac{9\Lambda^4}{4\pi^2} & \rightarrow & \Omega_0 \nonumber \\
\frac{1}{G} - \frac{\Lambda^2}{\pi^2} & \rightarrow & \frac{1}{G} \, .
\ea
For convenience we also define
\be
\Lambda^\prime \equiv 2^{2/3} e^{-1} \Lambda \, .
\ee
Then the thermodynamic potential can be written as
\be
\Omega = \Omega_0 - \frac{9\mu^4}{12\pi^2} + \frac{3 \Delta^2}{G}
- \frac{6}{\pi^2} \mu^2 \Delta^2 \ln \left( \frac{\Lambda^\prime}{\Delta}\right) 
\, .
\ee
The solutions to the gap equation, $\partial \Omega/\partial \Delta = 0$, are 
either $\Delta = 0$ or $\Delta = \Delta_0$ with the latter determined by
\be\frac{1}{G} = \frac{\mu^2}{\pi^2} \left[ 2 \ln \left( 
\frac{\Lambda^\prime}{\Delta_0}
\right) - 1 \right] \, .
\ee
The oft-used requirement that a change in $\Lambda^\prime$ should result in a 
change in $G$ such that $\Delta_0$ be left unchanged means that
\be
G = \frac{G_0}{1 + (2 \mu^2 G_0/\pi^2) \ln (\Lambda^\prime/\Lambda^\prime_0)}
\ee
So in this case we can trade the pair $G,\Lambda^\prime$ for one parameter, 
namely, $\Delta_0$.  Then we obtain the elegant formula
\be
\Omega = \Omega_0 - \frac{9\mu^4}{12\pi^2}+ \frac{3}{\pi^2} \mu^2 \Delta^2 
\left[ 2\ln \left( \frac{\Delta}{\Delta_0}\right) - 1 \right] \, .
\ee
This obviously has a minimum at $\Delta = \Delta_0$.

\subsection{CFL with $m_s \neq 0$}

Next we consider CFL with $\mu_S \neq 0$.  The thermodynamic potential now takes 
the form
\be
\Omega = \Omega_0 - \frac{3}{4\pi^2} \bar{\mu}^4
+ \frac{3\Delta^2}{\pi^2} \left[ 2 \mu^2 \ln \left( 
\frac{\Lambda^\prime}{\Delta_0}\right)
-2\bar{\mu}^2
\ln \left( \frac{\Lambda^\prime}{\Delta}\right) 
-\mu^2 \right] \, ,
\ee
where, as before $\bar{\mu} \equiv \mu - \oneth \mu_S$. The solution to the gap 
equation is either $\Delta = 0$ or the solution to the equation
\be
\bar{\mu}^2
\ln \left( \frac{\Delta}{\Lambda^\prime}\right)
=  \mu^2 \ln \left( \frac{\Delta_0}{\Lambda^\prime}\right)
+ \frac{1}{18} (6 \mu - \mu_S) \mu_S \, .
\label{CFLgapS}
\ee
When $\mu_S \neq 0$ the dependence on the cut-off no longer 
drops out.  In the 
limit that $\mu_S \ll \mu$ the solution to the above equation is approximately
\be
\Delta = \Delta_0 \left\{ 1 - \frac{\mu_S}{3\mu}
\left[ 2 \ln \left( \frac{\Lambda^\prime}{\Delta_0}\right)
- 1 \right] \right\} \, .
\label{gapS}
\ee
Hence the gap decreases linearly with $\mu_S/\mu$.

Label the solution to the gap equation (\ref{CFLgapS}) as 
$\Delta_{\rm CFL}(\mu_S)$.  This is a definite function of $\mu_S$.  In terms of 
it the thermodynamic potential may be written
\be
\Omega = \Omega_0 - \frac{3\bar{\mu}^4}{4\pi^2}+ \frac{3}{\pi^2} \bar{\mu}^2 
\Delta^2 \left[ 2\ln \left( \frac{\Delta}{\Delta_{\rm CFL}(\mu_S)}\right) - 1 
\right] \, .
\ee
This obviously has a minimum at $\Delta = \Delta_{\rm CFL}(\mu_S)$; the 
dependence on $\Lambda^\prime$ is implicit through the solution to 
(\ref{CFLgapS}).

Let us compare with Alford, Kouvaris, and Rajagopal \cite{alf05} who solved the 
gap equation numerically without approximation.  They used $\Delta_0 = 25$ MeV, 
$\mu = 500$ MeV, and $\Lambda = 800$ MeV ($\Lambda^\prime = 467.18$ MeV).  In 
the CFL phase the first gapless mode appears when $\mu_S = \Delta$.  From eq. 
(\ref{gapS}) we find that this occurs when $\Delta = \mu_S \approx 23.1$ MeV.  
This is consistent with their Fig. 1.

\subsection{Gapless CFL}

As $\mu_S$ increases, eventually one or more of the quark pairings will have a 
mismatch of chemical potentials that becomes as large as the gap.  This stress 
breaks apart some of the pairs and reduces that value of the gap.  Let us write 
the dispersion relation of quasi-quarks that first become gapless in the form
\be
E(p)= \sqrt{(p-\bar{\mu})^2+\Delta^2} - \delta\mu \, .
\label{ep}
\ee
The excitation energy for quasi-quarks becomes zero at the momenta
\be
p_{\pm}= \bar{\mu} \pm \sqrt{{\delta\mu}^2-\Delta^2} \, .
\label{pm}
\ee
This is appropriate when $\delta\mu \geq \Delta$.  Then there exist gapless 
modes in the blocking region $p\in(p_-,p_+)$. As a consequence, the gapless 
modes provide
\ba 
{\Omega}_g(\bar{\mu}, \delta \mu, \Delta) &=&
\int_{p_-}^{p_+} \frac{p^2 dp}{2\pi^2}
[\sqrt{(p-\bar{\mu})^2+\Delta^2} - \delta \mu] \nonumber \\
&=& \frac{\delta \mu}{24\pi^2} \sqrt{\delta \mu^2 - \Delta^2}
\left( 5 \Delta^2 - 2 \delta \mu^2 -12 \bar{\mu}^2 \right) \nonumber \\
&& +\frac{\Delta^2}{16\pi^2} \left( 4 \bar{\mu}^2 - \Delta^2 \right)
\ln \left( \frac{\delta \mu + \sqrt{\delta \mu^2 - \Delta^2}}
{\delta \mu - \sqrt{\delta \mu^2 - \Delta^2}} \right)
\label{pgap}
\ea
to the total free energy.  This must be added to (\ref{LargeL}).  At this point 
Alford, Kouvaris and Rajagopal solve the equations for gCFL numerically using 
the NJL model.  This is the gCFL phase.  The three chemical potentials, $\mu_e$, 
$\mu_3$ and $\mu_8$ evolve as a function of $\mu_S$ differently than in the CFL 
phase.  The three gaps are no longer equal but evolve such that $\Delta_I < 
\Delta_{II} < \Delta_{III}$ with $\Delta_I$ dropping the fastest, $\Delta_{II}$ 
lagging somewhat behind, and $\Delta_{III}$ not deviating much from $\Delta_0$.  
This is as expected.  At the onset of the gapless phase, channel III has no 
chemical potential mismatch and so the associated gap $\Delta_{III}$ should not 
change much.  The channels I and II both have a mismatch equal to $\mu_S$ and so 
their associated gaps should decrease; $\Delta_I$ decreases faster than 
$\Delta_{II}$ because the latter is retarded by the turn-on of the electric 
chemical potential.  We shall go no further into the gCFL since the phase with 
kaon condensation has lower free energy.

\subsection{CFL$K^0$}

Here we repeat the analysis of section 4.2 but with the chemical potentials and 
dispersion relations appropriate to the CFL$K^0$ phase. The thermodynamic 
potential is
\be
\Omega = \Omega_0 - \frac{3\bar{\mu}^2 }{4\pi^2} \left( \bar{\mu}^2
+ \frac{1}{6} \mu_S^2 \right)+ \frac{3\Delta^2}{\pi^2} \left[ 2 \mu^2 
\ln \left( \frac{\Lambda^\prime}{\Delta_0}\right)
-2\bar{\mu}^2
\ln \left( \frac{\Lambda^\prime}{\Delta}\right) 
-\mu^2 \right] \, ,
\ee
where, as before $\bar{\mu} \equiv \mu - \oneth \mu_S$. We have dropped terms of 
fourth order in $\Delta$ and $\mu_S$.  The functional dependence on the gap is 
identical to the CFL case.  Therefore the gap equation and its solutions are 
also identical.  We can thus write the thermodynamic potential in the form
\be
\Omega = \Omega_0 - \frac{3\bar{\mu}^2}{4\pi^2} \left( \bar{\mu}^2
+ \frac{1}{6} \mu_S^2 \right) + \frac{3}{\pi^2} \bar{\mu}^2 
\Delta^2 \left[ 2\ln \left( \frac{\Delta}{\Delta_{\rm CFL}(\mu_S)}\right) - 1 
\right] \, .
\ee
\subsection{Gapless CFL$K^0$}

Finally we turn our attention to the gCFL$K^0$ phase.  The difference in free 
energy between the CFL and CFL$K^0$ phases was already quoted in 
(\ref{deltaCFLK}).  We now face the problem of how to construct the free energy 
for the gCFL$K^0$ phase.  The above NJL model allows for different values of the 
gaps in the different channels, and it has the dynamics to determine these in 
terms of the coupling constant $G$ and the momentum cut-off $\Lambda$, but it 
does not allow for kaon condensation.  A NJL model was constructed by Forbes 
\cite{forbes} which does include kaon condensation.  (See also \cite{buba}.)  
The resulting calculations involve extensive numerical computation, but only the 
location of the onset of the gCFL$K^0$ phase was studied.  The effective field 
theory that describes kaon condensation assumes a common gap and that this gap 
is determined by other considerations.  In principle an effective field theory 
should be constructed that makes allowance for differing gaps; this is a 
formidable task that has yet to be worked out.  We are interested in describing 
the gCFL$K^0$ phase itself, at least for a small range of $\mu_S$, and not just 
the location of its onset.

We shall construct a thermodynamic potential from the following pieces.  The 
first piece is the NJL form (\ref{basicNJL}).  Practically we will use the large 
cut-off limit (\ref{LargeL}) and drop the terms of order $\Delta^4$.  The second 
piece is the contribution of the gapless modes described by (\ref{pgap}) which 
is, however, really a part of the NJL model.  The third and final piece is the 
kaon condensation free energy.  We should allow for the possibility that the 
values of the chemical potentials may be different once the gapless phase is 
entered.  We take this from Kryjevski \cite{kry03}.
\be
\Omega_{K^0} = - f_{\pi}^2 \left[ \thalf \mu_S^2
+ \thalf \left( \mu_3 + \thalf \mu_S - \mu_e \right)^2
+ \twoth \left( \mu_8 + \oneqt \mu_S  - \thalf \mu_e \right)^2 \right]
\ee 
Here we use the transcription
\bd
eA^0 \rightarrow \mu_e \, , \,\,\,\,
gA_3^0 \rightarrow - \mu_3 \, , \,\,\,\,
gA_8^0 \rightarrow - \frac{2}{\sqrt{3}} \mu_8 \, .
\ed
This contribution is normalized such that it reflects the free energy difference 
between the CFL$K^0$ and the CFL phases, namely,
\bd
\Omega_{K^0}(\mu_e=0, \mu_3=-\thalf \mu_S, \mu_8=-\oneqt \mu_S) = 
- \thalf f_{\pi}^2 \mu_S^2 \, .
\ed
Putting everything together, our model for the CFL$K^0$/gCFL$K^0$ matter is 
represented by the function
\ba
\Omega_{{\rm CFL}K^0} &=& \Omega^{\rm neutral}_{\rm unpaired}
 + \frac{3}{4\pi^2} \bar{\mu}^2 \mu_S^2 + \frac{\bar{\mu}^2}{\pi^2}
\sum_{i=I}^{III} \Delta_i^2 
\left[ 2\ln \left( \frac{\Delta_i}{\Delta_{\rm CFL}(\mu_S)}\right) - 1 \right]
\nonumber \\
&-& f_{\pi}^2 \left[ \thalf \mu_S^2+ \thalf \left( \mu_3 + \thalf \mu_S - \mu_e 
\right)^2
+ \twoth \left( \mu_8 + \oneqt \mu_S  - \thalf \mu_e \right)^2 \right]
 \nonumber \\
&+& \sum_{i=I}^{III}\theta(\delta \mu_i^2 - \Delta_i^2) \,
{\Omega}_g(\bar{\mu}_i, \delta \mu_i, \Delta_i) 
- \frac{\mu_e^4}{12\pi^2} \, .
\label{gCFLK}
\ea
The function $\Delta_{\rm CFL}(\mu_S)$ is the solution to eq. (\ref{CFLgapS}) 
whose approximate solution is given by eq. (\ref{gapS}).  The average and 
difference of chemical potentials in the $i$'th channel, $\bar{\mu}_i$ and 
$\delta \mu_i$, are the functions of $\mu_e$, $\mu_3$, $\mu_8$ to be found in 
Tables II and III.  The six variables $\mu_e$, $\mu_3$, $\mu_8$, $\Delta_I$, 
$\Delta_{II}$, and $\Delta_{III}$ must be determined on the basis of this 
function.  Note that the solutions give the physical thermodynamic potential of 
eq. (\ref{deltaCFLK}) when there are no gapless modes.  This function allows us 
to study the transition from CFL$K^0$ to gCFL$K^0$.  Note that the gapless mode 
in channel II is what will first drive the chemical potentials and the gap away 
from their values in the CFL$K^0$ phase.  Before proceeding we will make one 
other useful approximation; we set $\bar{\mu}_i = \bar{\mu}$.  This has a 
minimal effect because it is the difference $\delta \mu_i$ which drives the 
gapless phase, not the average.  In addition we use $\bar{\mu}$ in place of 
$\mu$ when evaluating $f_{\pi}$.  Although $f_{\pi}$ has not been determined 
when the quark masses are nonzero, this replacement has minimal numerical 
impact.

Consider what happens as $\mu_S$ increases, or equivalently what happens as the 
baryon density decreases.  Prior to the onset of the gapless modes, we have 
CFL$K^0$ matter.  The chemical potentials are $\mu_e=0, \mu_3=-\thalf \mu_S, 
\mu_8=-\oneqt \mu_S$.  All three gaps are equal, $\Delta_I = \Delta_{II} = 
\Delta_{III} = \Delta_{\rm CFL}(\mu_S)$, and decrease with increasing $\mu_S$.  
The thermodynamic potential is given by
\be
\Omega^{\rm neutral}_{{\rm CFL}K^0} = \Omega^{\rm neutral}_{\rm unpaired}
+ \frac{3\bar{\mu}^2}{4 \pi^2} \left( \mu_S^2 - 4 \Delta^2_{\rm CFL} \right) 
- \frac{1}{2} f_{\pi}^2 \mu_S^2 \, .
\ee

The channel II goes gapless when $\mu_S = \textstyle{\frac{4}{3}} 
\Delta_{\rm CFL}(\mu_S) \approx 1.206 \Delta_0$.  The difference between the 
factor of 4/3 (first obtained by Kryjevski and Yamada \cite{kry2}) and 1.206 
(very close to the value obtained by Forbes \cite{forbes}) is due to the 
decrease in the gap with increasing $\mu_S$.  To study this phase we need to 
solve the equations
\be 
\partial \Omega_{{\rm CFL}K^0}/\partial \mu_3 =
\partial \Omega_{{\rm CFL}K^0}/\partial \mu_8 = 
\partial \Omega_{{\rm CFL}K^0}/\partial \mu_e = 0
\ee 
to obtain the chemical potentials.  Allowing for the possibility that any of the 
channels may be gapless, these can be recast in the form
\be
f^2_{\pi} \left[ 2\mu_S -4\mu_e + 3\mu_3 + 2\mu_8 \right]
= \frac{3}{2} \frac{\partial \Omega_g}{\partial \delta \mu_{II}}
+ \frac{3}{2} \frac{\partial \Omega_g}{\partial \delta \mu_{III}}
+ \frac{\mu_e^3}{\pi^2}
\label{cheme}
\ee
\be
f^2_{\pi} \left[ \mu_S -2\mu_e + 2\mu_3 \right]
= \frac{1}{2} \frac{\partial \Omega_g}{\partial \delta \mu_{I}}
-\frac{1}{2} \frac{\partial \Omega_g}{\partial \delta \mu_{II}}
- \frac{\partial \Omega_g}{\partial \delta \mu_{III}}
\label{chem3}
\ee
\be
f^2_{\pi} \left[ \mu_S -2\mu_e + 4\mu_8 \right]
= -\frac{3}{2} \frac{\partial \Omega_g}{\partial \delta \mu_{I}}
- \frac{3}{2} \frac{\partial \Omega_g}{\partial \delta \mu_{II}}
\label{chem8}
\ee
Here we have calculated the derivatives $\partial \delta \mu_i/\partial 
\mu_{\alpha}$ (where $\alpha = e, 3,8$) by using 
Table II.  

When entering the gapless region it is sometimes useful to employ the limiting 
form of $\Omega_g$ when $\sqrt{\delta \mu^2 - \Delta^2} \ll \delta 
\mu, \, \Delta \ll \bar{\mu}$, specifically
\ba
\Omega_g & \approx & - \frac{\bar{\mu}^2}{2\pi^2 \delta \mu}
\left( \delta \mu^2 - \Delta^2 \right)^{3/2}  \nonumber \\
\frac{\partial \Omega_g}{\partial \delta \mu} & \approx &
- \frac{3\bar{\mu}^2}{2\pi^2}
\left( \delta \mu^2 - \Delta^2 \right)^{1/2} \nonumber \\
\frac{\partial \Omega_g}{\partial \Delta} & \approx &
\frac{3\bar{\mu}^2 \Delta}{2\pi^2 \delta \mu}
\left( \delta \mu^2 - \Delta^2 \right)^{1/2} \, .
\label{approxOg}
\ea
We shall henceforth use this expression.  Its validity will be checked {\it a 
posteriori}.  Using this approximation, the gap in channel $i$ must 
satisfy either $\Delta_i = 0$ or the equation
\be
\ln \left( \frac{\Delta_i}{\Delta_{\rm CFL}} \right) +
\frac{3}{8} \frac{\sqrt{\delta \mu_i^2 - \Delta_i^2}}
{\delta \mu_i} \theta\left(\delta \mu_i^2 - \Delta_i^2 \right) = 0 \, .
\ee

We are left with solving the above set of six equations to find the behavior of 
$\mu_e$, $\mu_3$, $\mu_8$, $\Delta_I$, $\Delta_{II}$ and $\Delta_{III}$ as 
functions of $\mu_S$.  At this point it is already clear that $\mu_e$ will 
increase from 0 and a gap will decrease from $\Delta_{\rm CFL}(\mu_S)$ in a 
closely coupled way once we enter a gapless region.

The electron chemical potential $\mu_e$ will become nonzero once $\ttqt \mu_S$ 
exceeds $\Delta_{\rm CFL}$.  Therefore it is sensible to look for a power series 
expansion of the unknowns in terms of $x \equiv \ttqt \mu_S - \Delta_{\rm CFL}$.  
The results are
\ba
\mu_e &=& x +
\frac{x^3}{12}\left( \frac{1}{\bar{\mu}^2} - 
\frac{1}{\pi^2 f_{\pi}^2} \right)
+ \cdot \cdot \cdot \nonumber \\
\mu_3 &=& -\thalf \mu_S + x +
\frac{x^3}{12\bar{\mu}^2}
+ \cdot \cdot \cdot \nonumber \\
2\mu_8 &=& -\thalf \mu_S + x +
\frac{x^3}{12}
\left( \frac{1}{\bar{\mu}^2} + \frac{2}{\pi^2 f_{\pi}^2} \right)
+ \cdot \cdot \cdot \nonumber \\
\Delta_I &=& \Delta_{\rm CFL} \nonumber \\
\Delta_{II} &=& \Delta_{\rm CFL} -
\frac{x^3}{12\bar{\mu}^2}
+ \cdot \cdot \cdot \nonumber \\
\Delta_{III} &=& \Delta_{\rm CFL} \nonumber \\
\delta \mu_I &=& \thalf \mu_S  
- \frac{x^3}
{24\pi^2 f_{\pi}^2}
+ \cdot \cdot \cdot \nonumber \\
\delta \mu_{II} &=& \Delta_{\rm CFL} -
\frac{x^3}{12\bar{\mu}^2}
+ \cdot \cdot \cdot \nonumber \\
\delta \mu_{III} &=& \Delta_{\rm CFL} - \thalf \mu_S
- \frac{x^3}{12}
\left( \frac{1}{\bar{\mu}^2} - \frac{1}{2\pi^2 f_{\pi}^2} \right)
+ \cdot \cdot \cdot 
\label{powersol1}
\ea
The gaps $\Delta_I = \Delta_{III} = \Delta_{\rm CFL}(\mu_S)$ are undeviated from 
the CFL$K^0$ phase.  The relative order of magnitude of the corrections 
displayed above, before the ellipses, with $x \approx 10$ MeV and 
$\mu \approx 500$ MeV, is $10^{-4}$, indicating that this is 
a very good expansion.  This also supports the validity of using the 
approximation (\ref{approxOg}).  To the order displayed in eq. (\ref{powersol1}) 
the important quantity $\sqrt{\delta \mu_{II}^2-\Delta_{II}^2}$ vanishes.  It 
can be determined to lowest nonvanishing order to be
\be
\sqrt{\delta\mu_{II}^2-\Delta_{II}^2} = \frac{2}{9} 
\frac{x^3}{\bar{\mu}^2}+ \cdot \cdot \cdot
\ee
by inserting the expressions for the chemical potentials from (\ref{powersol1}) 
into (\ref{chem3}) or (\ref{chem8}).

The behavior of $\mu_3$, $\mu_8$ and $\mu_e$ as functions of $\mu_S$ are shown 
in Fig. 1 while the behavior of $\Delta_I=\Delta_{III}$ and $\Delta_{II}$ are 
shown in Fig. 2.  To obtain these figures we used eqs. (\ref{CFLgapS}) and 
(\ref{powersol1}).  We used the same numerical values for the parameters as in 
section 4.2, namely, $\Delta_0 = 25$ MeV, $\mu = 500$ MeV, and $\Lambda = 800$ 
MeV.  Essentially all of the functions plotted are linear in $\mu_S$ as is 
apparent from the figures.  For $\mu_S < {\textstyle{\frac{4}{3}}} 
\Delta_{\rm CFL} \approx 1.206 \Delta_0 \approx 30.143$ MeV all three channels 
are fully gapped.  For larger values of $\mu_S$ channel II becomes gapless.  

Channel I becomes gapless when $\delta \mu_I = \Delta_{\rm CFL}$, which occurs 
when
\bd
\Delta_{\rm CFL} = \thalf \mu_S  
- \frac{\mu_S^3}{1536\pi^2 f_{\pi}^2}+ 
\cdot \cdot \cdot \nonumber \\
\ed
The chemical potentials and gaps acquire shifts proportional to 
\be
y = \frac{ \oneqt \left( \mu_S - x^3/12 \pi^2 f_{\pi}^2 \right)^2
-\Delta_{\rm CFL}^2}{ \thalf \left( \mu_S - x^3/12 \pi^2 f_{\pi}^2 \right) 
- (\pi^2 f_{\pi}^2/\bar{\mu}^2) \Delta_{\rm CFL}} \, .
\ee
Starting from 
(\ref{powersol1}) they are shifted as follows:
\ba
\mu_e & \rightarrow & \mu_e - \oneqt y +
\cdot \cdot \cdot \nonumber \\
\mu_3 & \rightarrow & \mu_3 - \ttqt y +
\cdot \cdot \cdot \nonumber \\
2\mu_8 & \rightarrow & 2\mu_8 + {\textstyle{\frac{5}{4}}} y
 + \cdot \cdot \cdot \nonumber \\
\Delta_I & \rightarrow & \Delta_I - 
\frac{\pi^2 f_{\pi}^2}{2 \bar{\mu}^2} y +
\cdot \cdot \cdot \nonumber \\
\Delta_{II} & \rightarrow & \Delta_{II} 
-\frac{x^2}{16 \bar{\mu}^2} y + \cdot \cdot \cdot \nonumber \\
\Delta_{III} & \rightarrow & \Delta_{III} + \cdot \cdot \cdot \nonumber \\
\delta \mu_I & \rightarrow & \delta \mu_I - \thalf y +
\cdot \cdot \cdot \nonumber \\
\delta \mu_{II} & \rightarrow & \delta \mu_{II}
+\cdot \cdot \cdot \nonumber \\
\delta \mu_{III} & \rightarrow & \delta \mu_{III}
 + \thalf y + \cdot \cdot \cdot \, .
\label{powersol2}
\ea
To lowest nonvanishing order
\be
\sqrt{\delta\mu_{I}^2-\Delta_{I}^2} = \frac{4}{3} 
\frac{\pi^2 f_{\pi}^2}{\bar{\mu}^2} y
\ee
Unlike the case of gapless channel II, the corrections here in channel I are not 
suppressed by a factor of $1/\bar{\mu}^2$.  They begin to deviate rather 
substantially from $\Delta_{\rm CFL}$.  However, we do not need to go very 
deeply into the gapless channel I phase before the color superconducting matter 
becomes unstable to complete breakup into unpaired quark matter. 

Channel I becomes gapless when $\mu_S > 2 \Delta_{\rm CFL} \approx 1.728 
\Delta_0 \approx 43.2$ MeV.  In the region where only channel II is gapless, 
$\Delta_{II} < \Delta_I = \Delta_{III} = \Delta_{\rm CFL}$, but not by much; it 
is not visible on the Fig. 2.  In the region where both I and II are gapless, 
$\Delta_{I}$ drops much faster than $\Delta_{II}$.

With the parameters we have used, channel II becomes gapless when the strange 
quark mass $m_s = \sqrt{2\mu\mu_S} > 174$ MeV, and channel I becomes gapless 
when $m_s > 208$ MeV.

The thermodynamic potential, as measured relative to the thermodynamic potential 
of neutral unpaired quark matter, in units of $3\mu^2\Delta_0^2/\pi^2$, is shown 
in Fig. 3.  There is a smooth transition as first channel II and then channel I 
become gapless.  Paired quark matter becomes unstable to breakup into unpaired 
quark matter when $\mu_S > 50.47$ MeV.  Coincidentally, this is approximately 
equal to $2\Delta_0$, the value one would get by ignoring the evolution of the 
gaps with $\mu_S$, the chemical potentials enforcing electric and color 
neutrality, and the formation of gapless modes.

\section{Chromomagnetic Instabilities}

Finally, we analyze qualitatively the possibility of eliminating the
chromomagnetic instability that occurs in gCFL. In the
gCFL case, channels I and II become gapless simultaneously at the
point $(\mu_e,\mu_S)=(0,\Delta_{\rm CFL})$.  The fact that the $gs-bd$ and
$rs-bu$ pairings become breached simultaneously has an important consequence. As 
stressed in \cite{cas05,fuku}, this feature is responsible for the
instability which occurs for $A_1$ and $A_2$ gluons.  Since the self-energy for 
$A_1$ and $A_2$ stems from the loop diagram composed of $gs$ and $rs$ quarks, 
the coexistence of gapless (unpaired) $gs$ and $rs$ modes provides a singular
contribution to the corresponding Meissner masses. In the gCFL$K^0$
case, however, the gapless formation in channel I is delayed relative to 
channel II.  Therefore, singularities should no longer exist in the
self-energy for $A_{1,2}$ Meissner masses, and the corresponding
instability should disappear in the gCFL$K^0$ phase.

On the other hand, the feature that the gCFL location is very close to its 
critical line in the $\mu_e - \mu_S$ plane has been pointed out to be primarily 
responsible for the instability of $A_3$ and $A_8$ gluons and $A_\gamma$ photons  
\cite{fuku}. For instance, the (almost) infinitesimal channel II
gapless strength in gCFL leads to the singularities in the
$A_{3,8,\gamma}$ Meissner masses \cite{fuku}. In the gCFL$K^0$ case, however,
the location is not limited to the vicinity of the critical line.
Instead, a finite region in the $\mu_e - \mu_S$ plane is available
for the gCFL$K^0$ existence. Thus such kind of instability might no
longer appear, at least for some region in the $\mu_e - \mu_S$ plane.

Since only $A_{1,2}$ and $A_{3,8,\gamma}$ exhibit imaginary Meissner
masses in gCFL, the previous-predicted instability is very likely to
be removed in the gCFL$K^0$ phase. Physically, it can be understood
from the viewpoint that the chromomagnetic instability is
eliminated by building on the proper vacuum. It has been suggested
that nonzero vacuum expectation values of gluons such as $\langle
A^0_3\rangle$ and/or $\langle A^0_8\rangle$ are helpful for removing
the gCFL instability \cite{cas05}. In \cite{gorb}, the instability in two-flavor 
superconductor was argued to be resolved by gluon condensation.
In the present work, the nonzero $\langle A^0_3\rangle$ and $\langle
A^0_8\rangle$ (anisotropic vacuum) derived from the kaon
condensation have been attributed to the change of the
electric/color neutral solution and then been considered in the
study of the gapless formation. In this sense, the conjecture that gCFL$K^0$ is 
not chromomagnetically unstable is not very surprising. However, intrinsic links 
between the $p$-wave kaon supercurrent phase establised in \cite{sch06} and the
presently discussed gCFL$K^0$ phase still remain to be clarified.
Also, we cannot rule out the possibility that other instabilities, especially 
for $A_4$, $A_5$, $A_6$ and $A_7$ gluons, arise once again. The quantitative 
calculation of the gluonic Meissner masses in gCFL$K^0$ is still necessary, but 
this is beyond the scope of the present work.

\section{Conclusion}

In summary, we investigated electric/color neutrality and gapless
formation in the CFL matter with $K^0$ condensation. By taking the
CFL$K^0$ neutral solution into account, we clarify why the gCFL$K^0$
formation is delayed in comparison with the gCFL case. More importantly, it is 
found that the gapless phenomenon for down-strange pairing (channel I) is
delayed while that for up-strange pairing (channel II) occurs first. The novel 
phase structure implies that the previously predicted gCFL instability might
be removed or at least weakened. These conclusions are likely to be important 
for fully understanding the unconventional CFL phases in the presence of strange 
quarks with nonzero masses $m_S$ and eletrons $\mu_e$.

Of course, there are still some unanswered aspects in the present work. First of 
all, the $K^0$ condensed CFL matter has been treated as a background for the 
gapless formation, which is safe only if $\mu$ is not too small; otherwise, the 
condensation may be supressed by instanton effect \cite{ins}. The gCFL$K^0$ 
phase is lower in free energy than the gCFL phase.  Thus, our suggested gapless 
phase should replace the gCFL in neutron star cores \cite{alf05b}, ignoring 
any possible gCFL$K^0$ instability.  Secondly, more physics involving 
Goldstone-mode condensation should be addressed. These include the possibilities 
of charged kaon and other meson condensations. For instance, if $\mu_e$ is large 
enough and $K^-$ condensation occurs, the gapless phenomenon for up-down pairing 
in channel III needs to be included.  Some of the above-mentioned problems are 
being investigated.

\section*{Acknowledgements}

We thank M. Alford, A. Kryjevski, K. Rajagopal and T. Sch\"afer for helpful 
communication during the course of this work.  This work was supported by 
National Natural Science Foundation of China (NSFC) grant 10405012 and by the US 
Department of Energy (DOE) under grant DE-FG02-87ER40328.

\section*{Appendix}

An integral that appears frequently is
\bd
\int_0^{\Lambda} dp \, p^2 \sqrt{(p-\bar{\mu})^2 + \Delta^2} =
\frac{1}{8} \bar{\mu} \left[ 6 \bar{\mu}^2 + \Delta^2 \right] 
\sqrt{\bar{\mu}^2 + \Delta^2}
\ed
\bd
\frac{1}{8} \left( \Lambda - \bar{\mu} \right) 
\left[ 2 \left( \Lambda - \bar{\mu} \right)^2 + 4 \bar{\mu}^2 + \Delta^2 \right]
\sqrt{ \left( \Lambda - \bar{\mu} \right)^2 + \Delta^2}
\ed
\bd
+ \frac{2}{3} \left\{ \left[\left( \Lambda - \bar{\mu} \right)^2 + \Delta^2 
\right]^{3/2} - \left[ \bar{\mu}^2 + \Delta^2 \right]^{3/2} \right\}
\ed
\bd
+ \frac{1}{8} \Delta^2 \left( 4 \bar{\mu}^2 - \Delta^2 \right)
\ln \left( \frac{ \Lambda - \bar{\mu} + \sqrt{ \left( \Lambda - \bar{\mu} 
\right)^2 + \Delta^2}}{\sqrt{\bar{\mu}^2 + \Delta^2} - \bar{\mu}} \right)
\ed
The limit of this expression as $\Lambda \rightarrow \infty$ is
\bd
\int_0^{\Lambda} dp \, p^2 \sqrt{(p-\bar{\mu})^2 + \Delta^2}
\rightarrow \frac{1}{4} \Lambda^4 - \frac{1}{3} \bar{\mu} \Lambda^3
+ \frac{1}{4} \Delta^2 \Lambda^2 + \frac{1}{2} \bar{\mu} \Delta^2 \Lambda
\ed
\bd
+ \frac{1}{8} \Delta^2 \left( 4 \bar{\mu}^2 - \Delta^2 \right)
\ln \left( \frac{2 \Lambda}{\sqrt{\bar{\mu}^2 + \Delta^2} - \bar{\mu}} \right)
+ \frac{1}{12} \bar{\mu}^4 - \frac{1}{2} \bar{\mu}^2 \Delta^2 + 
\frac{1}{32} \Delta^4
\ed
\bd
+ \frac{1}{24} \bar{\mu} \left( 2 \bar{\mu}^2 - 13 \Delta^2 \right)
\sqrt{\bar{\mu}^2 + \Delta^2}
\ed
The terms which are odd in $\bar{\mu}$ cancel when particles and anti-particles 
are added together.

\newpage
\renewcommand{\arraystretch}{1.5}

\noindent
Table I: Effective quark chemical potentials.  The third and fourth columns 
evaluate them in the electron-free $\left(\mu_e=0\right)$ CFL 
$\left(\mu_3=0, \mu_8=-\mu_S\right)$ and CFLK$^0$ $\left(\mu_3= - 
\frac{1}{2}\mu_S, \mu_8= - \frac{1}{4}\mu_S\right)$ phases.\\
\begin{center}
\begin{tabular}{|c||c|c|c|}
\hline
\hspace*{10mm}& effective chemical potential & CFL & CFLK$^0$ \\
\hline
\hline
$ru$ & $\mu-\frac{2}{3}\mu_e+\frac{1}{2}\mu_3+\frac{1}{3}\mu_8$
& $\mu-\frac{1}{3}\mu_S$ & $\mu-\frac{1}{3}\mu_S$ \\
\hline
$gd$ & $\mu+\frac{1}{3}\mu_e-\frac{1}{2}\mu_3+\frac{1}{3}\mu_8$
& $\mu-\frac{1}{3}\mu_S$ & $\mu+\frac{1}{6}\mu_S$ \\
\hline
$bs$ & $\mu+\frac{1}{3}\mu_e-\frac{2}{3}\mu_8-\mu_S$
& $\mu-\frac{1}{3}\mu_S$ & $\mu-\frac{5}{6}\mu_S$ \\
\hline
$rd$ & $\mu+\frac{1}{3}\mu_e+\frac{1}{2}\mu_3+\frac{1}{3}\mu_8$ &
$\mu-\frac{1}{3}\mu_S$ & $\mu-\frac{1}{3}\mu_S$ \\
\hline
$gu$ & $\mu-\frac{2}{3}\mu_e-\frac{1}{2}\mu_3+\frac{1}{3}\mu_8$ &
$\mu-\frac{1}{3}\mu_S$ & $\mu+\frac{1}{6}\mu_S$ \\
\hline
$rs$ & $\mu+\frac{1}{3}\mu_e+\frac{1}{2}\mu_3+\frac{1}{3}\mu_8
-\mu_S$ & $\mu-\frac{4}{3}\mu_S$ & $\mu-\frac{4}{3}\mu_S$ \\
\hline
$bu$ & $\mu-\frac{2}{3}\mu_e-\frac{2}{3}\mu_8$ &$\mu+\frac{2}{3}\mu_S$ & 
$\mu+\frac{1}{6}\mu_S$ \\
\hline
$gs$ & $\mu+\frac{1}{3}\mu_e-\frac{1}{2}\mu_3+\frac{1}{3}\mu_8
-\mu_S$ & $\mu-\frac{4}{3}\mu_S$ & $\mu-\frac{5}{6}\mu_S$ \\\hline
$bd$ & $\mu+\frac{1}{3}\mu_e-\frac{2}{3}\mu_8$ &
$\mu+\frac{2}{3}\mu_S$ & $\mu+\frac{1}{6}\mu_S$ \\
\hline
\end{tabular}
\end{center}

\newpage

\noindent
Table II: Differences of effective quark chemical potentials relevant to 
pairing, defined as $\delta \mu = \frac{1}{2} \left(\mu_{q1}-\mu_{q2}\right)$ 
for the pair $q1$-$q2$.  The third and fourth columns evaluate them in the 
electron-free $\left(\mu_e=0\right)$ CFL 
$\left(\mu_3=0, \mu_8=-\mu_S\right)$ and CFL$K^0$ $\left(\mu_3=-
\frac{1}{2}\mu_S, \mu_8=-\frac{1}{4}\mu_S\right)$ phases.\\

\begin{center}
\begin{tabular}{|c||c|c|c|}
\hline
\hspace*{10mm}& $\delta \mu$ & CFL & CFLK$^0$ \\
\hline
\hline
$ru-gd$ & $\frac{1}{2}\left(-\mu_e+\mu_3\right)$
& 0 & $-\frac{1}{4}\mu_S$ \\
\hline
$ru-bs$ & $\frac{1}{2}\left(-\mu_e+\frac{1}{2}\mu_3+\mu_8+\mu_S\right)$
& 0 & $\frac{1}{4}\mu_S$ \\
\hline
$gd-bs$ & $\frac{1}{2}\left(-\frac{1}{2}\mu_3+\mu_8+\mu_S\right)$
& 0 & $\frac{1}{2}\mu_S$ \\
\hline
$bd-gs$ & $\frac{1}{2}\left(\frac{1}{2}\mu_3-\mu_8+\mu_S\right)$ &
$\mu_S$ & $\frac{1}{2}\mu_S$ \\
\hline
$bu-rs$ & $\frac{1}{2}\left(-\mu_e-\frac{1}{2}\mu_3-\mu_8+\mu_S\right)$ &
$\mu_S$ & $\frac{3}{4}\mu_S$ \\
\hline
$gu-rd$ & $\frac{1}{2}\left(-\mu_e-\mu_3\right)$ & 0 & $\frac{1}{4}\mu_S$ \\
\hline
\end{tabular}
\end{center}

\vspace*{5mm}
\noindent
Table III: Averages of effective quark chemical potentials relevant to 
pairing.  The third and fourth columns evaluate them in the 
electron-free $\left(\mu_e=0\right)$ CFL 
$\left(\mu_3=0, \mu_8=-\mu_S\right)$ and CFL$K^0$ $\left(\mu_3=-
\frac{1}{2}\mu_S, \mu_8=-\frac{1}{4}\mu_S\right)$ phases.\\

\begin{center}
\begin{tabular}{|c||c|c|c|}
\hline
\hspace*{10mm}& average & CFL & CFLK$^0$ \\\hline
\hline
$ru-gd$ & $\mu-\frac{1}{6}\mu_e+\frac{1}{3}\mu_8$
& $\mu-\frac{1}{3}\mu_S$ & $\mu-\frac{1}{12}\mu_S$ \\
\hline
$ru-bs$ & $\mu-\frac{1}{6}\mu_e+\frac{1}{4}\mu_3
-\frac{1}{6}\mu_8-\frac{1}{2}\mu_S$
& $\mu-\frac{1}{3}\mu_S$ & $\mu-\frac{7}{12}\mu_S$ \\
\hline
$gd-bs$ & $\mu+\frac{1}{3}\mu_e-\frac{1}{4}\mu_3
-\frac{1}{6}\mu_8-\frac{1}{2}\mu_S$& $\mu-\frac{1}{3}\mu_S$ & $\mu-
\frac{1}{3}\mu_S$ \\
\hline
$bd-gs$ & $\mu+\frac{1}{3}\mu_e-\frac{1}{4}\mu_3
-\frac{1}{6}\mu_8-\frac{1}{2}\mu_S$ &
$\mu-\frac{1}{3}\mu_S$ & $\mu-\frac{1}{3}\mu_S$ \\\hline
$bu-rs$ & $\mu-\frac{1}{6}\mu_e+\frac{1}{4}\mu_3
-\frac{1}{6}\mu_8-\frac{1}{2}\mu_S$ &
$\mu-\frac{1}{3}\mu_S$ & $\mu-\frac{7}{12}\mu_S$ \\
\hline
$gu-rd$ & $\mu-\frac{1}{6}\mu_e+\frac{1}{3}\mu_8$ & $\mu-\frac{1}{3}\mu_S$ & 
$\mu-\frac{1}{12}\mu_S$ \\
\hline
\end{tabular}
\end{center}

\newpage

\noindent
Table IV: Correspondence between quark and baryon descriptions with the 
condensate form ${\langle q^\alpha_i C\gamma_5q^\beta_j\rangle} \sim \Delta_I 
\epsilon^{\alpha\beta
1}\epsilon_{ij1} + \Delta_{II} \epsilon^{\alpha\beta 2}\epsilon_{ij2} 
+\Delta_{III} 
\epsilon^{\alpha\beta 3}\epsilon_{ij3}$.\\
\begin{center}
\begin{tabular}{|c||c|}
\hline
\hspace*{10mm}& baryon label \\
\hline
\hline
$ru$ & mixture of $\Sigma^0$, $\Lambda^0$, $\Lambda^8$ \\
\hline
$gd$ & mixture of $\Sigma^0$, $\Lambda^0$, $\Lambda^8$ \\\hline
$bs$ & mixture of $\Sigma^0$, $\Lambda^0$, $\Lambda^8$ \\
\hline
\hline
$gu$ & $\Sigma^+$ \\
\hline
$rs$ & $\Xi^-$ \\
\hline
$bu$ & $p$ \\
\hline
$gs$ & $\Xi^0$ \\
\hline
$bd$ & $n$ \\
\hline
\end{tabular}

\end{center}

\newpage

\begin{figure}
\begin{center}\includegraphics[width=5.0in,angle=90]{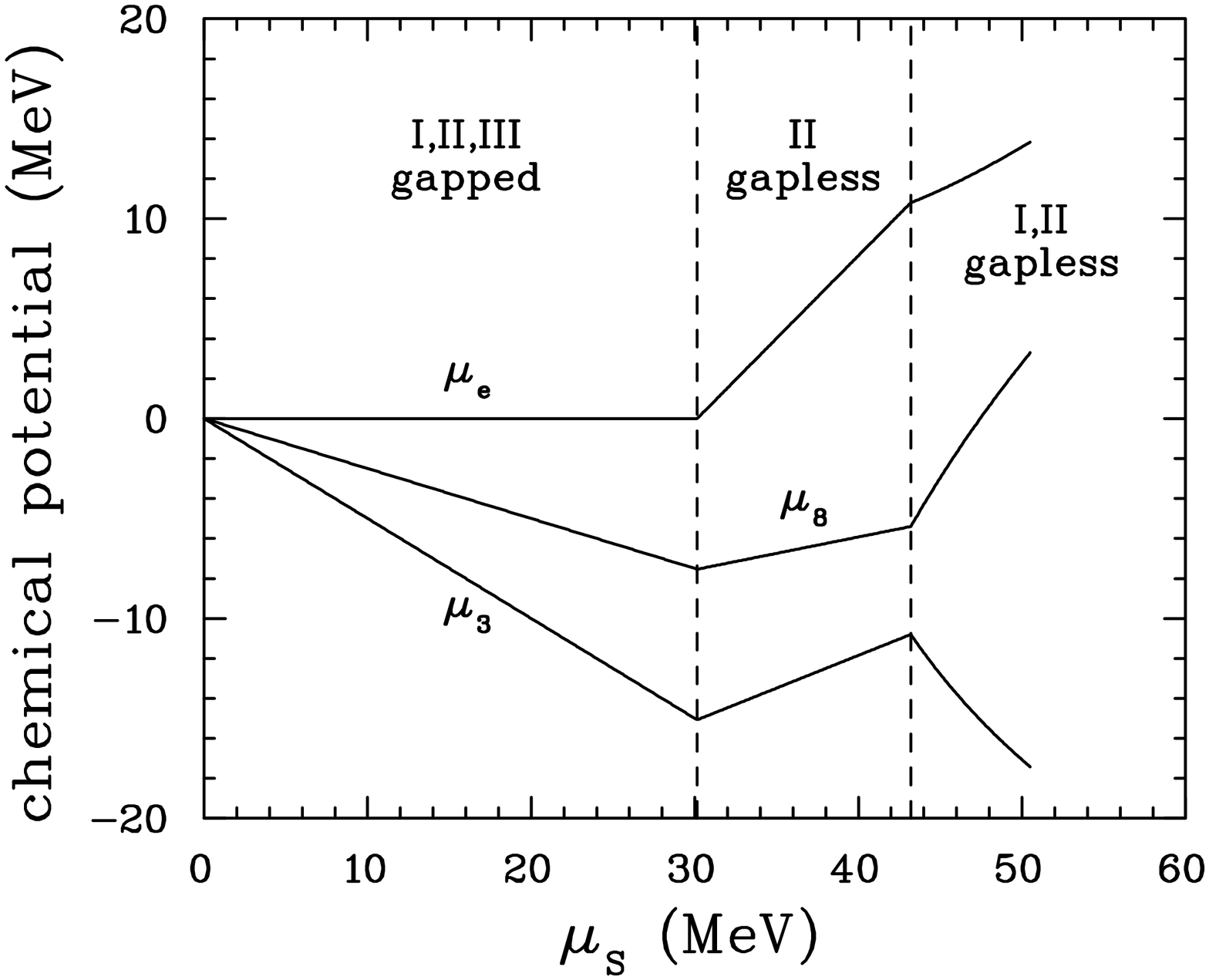}
\caption{Chemical potentials as functions of $\mu_S$.  The numerical parameters 
chosen are $\mu = 500$ MeV, $\Delta_0 = 25$ MeV, and $\Lambda = 800$ MeV.  The 
curves are terminated when the paired phase becomes unstable to decay to the 
unpaired phase.}
\end{center}
\end{figure}

\begin{figure}
\begin{center}\includegraphics[width=5.0in,angle=90]{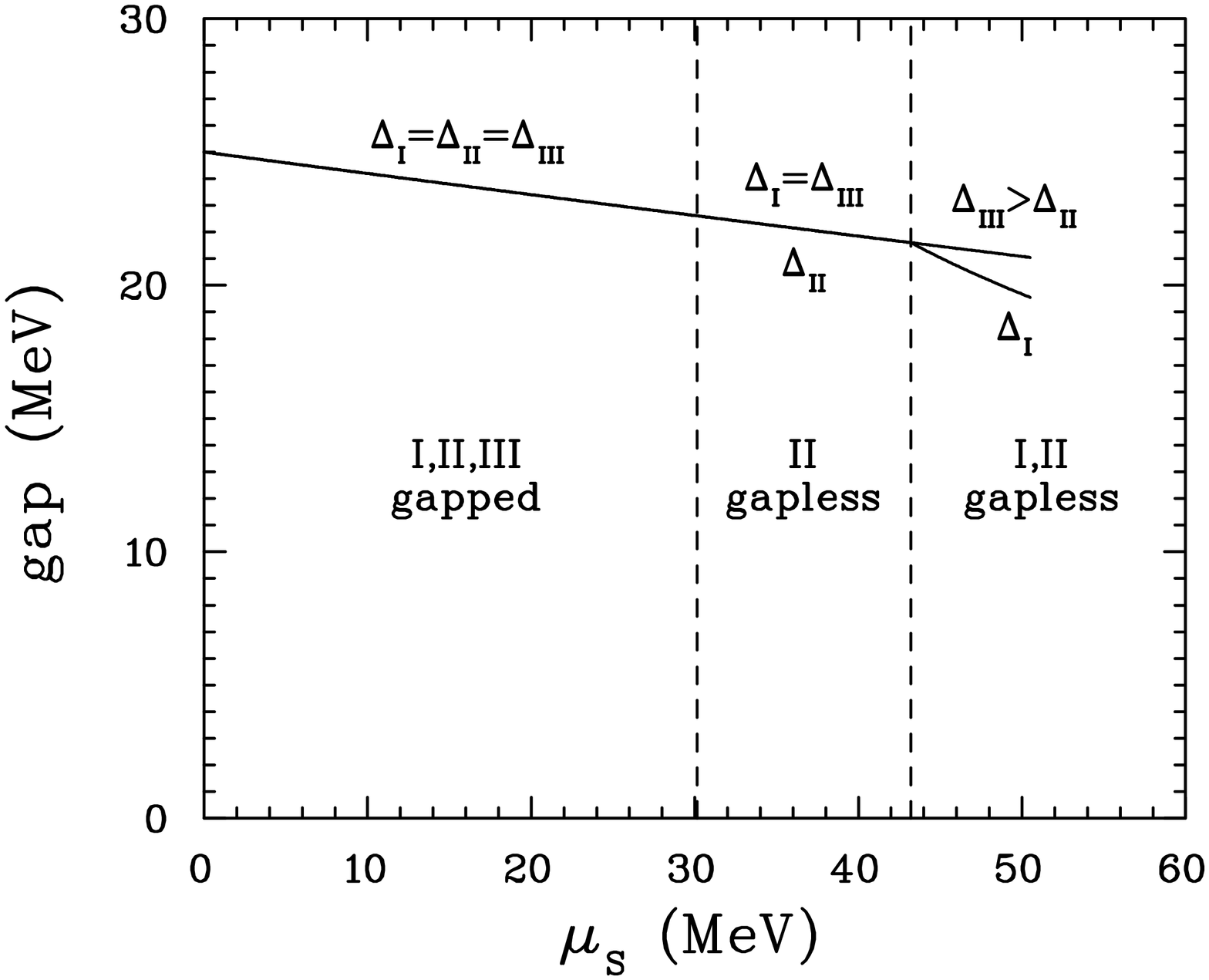}
\caption{The gaps as functions of $\mu_S$.  The numerical parameters chosen are 
$\mu = 500$ MeV, $\Delta_0 = 25$ MeV, and $\Lambda = 800$ MeV.  The curves are 
terminated when the paired phase becomes unstable to decay to the unpaired 
phase.}
\end{center}
\end{figure}

\begin{figure}
\begin{center}\includegraphics[width=5.0in,angle=90]{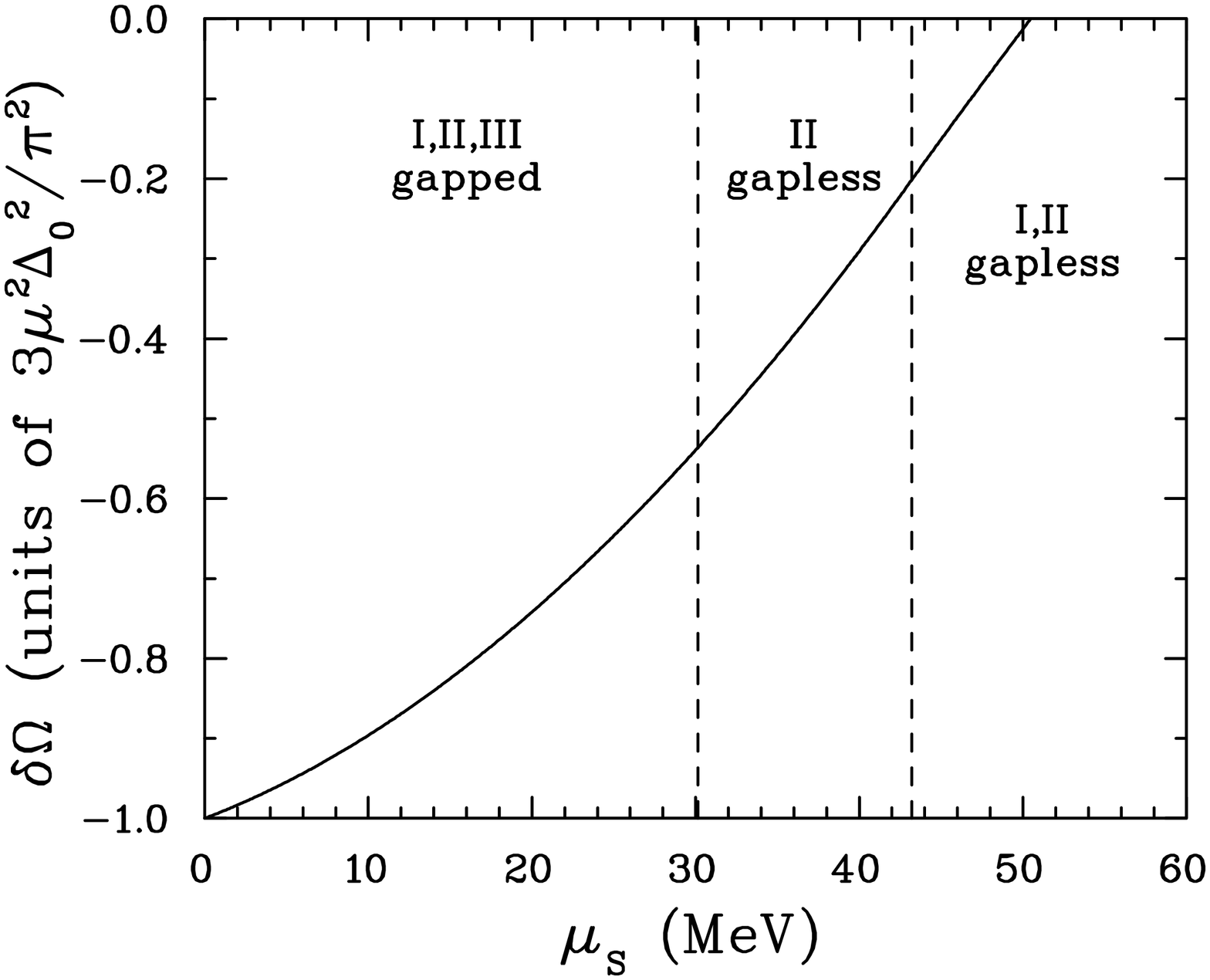}
\caption{The difference in thermodynamic potentials between the paired and 
unpaired phases as a function of $\mu_S$.  The numerical parameters chosen are 
$\mu = 500$ MeV, $\Delta_0 = 25$ MeV, and $\Lambda = 800$ MeV.  The transition 
occurs when $\mu_S \approx 50.47$ MeV.}
\end{center}
\end{figure}


\begin{thebibliography}{99}

\bibitem{alf}
 M. Alford, K. Rajagopal and F. Wilczek, Phys. Lett. {\bf B 422}, 247 (1998);
 M. Alford, K. Rajagopal and F. Wilczek, Nucl. Phys. {\bf B 537}, 443 (1999).

\bibitem{alf03}For reviews, see K. Rajagopal and F. Wilczek, hep-ph/0011333; 
T.Sch\"{a}fer, hep-ph/0304281; M. Alford, Prog. Thero. Phys. Suppl.
{\bf 153}, 1 (2004); D. H. Rischke, Prog. Part. Nucl. Phys. 
{\bf 52}, 197 (2004).

\bibitem{liu}   
E. Gubankova, W. V. Liu and F. Wilczek, Phys. Rev. Lett. {\bf
91}, 032001 (2003).

\bibitem{alf04}
 M. Alford, C. Kouvaris and K. Rajagopal, Phys. Rev. Lett. {\bf 92}, 222001
 (2004).
\bibitem{alf05} M. Alford, C. Kouvaris and K. Rajagopal, Phys. Rev. D
{\bf 71}, 054009 (2005).

\bibitem{huang} I. Shovkovy and M. Huang, Phys. Lett. {\bf B 564},
205 (2003).

\bibitem{cas05}
R. Casalbuoni, D. Gatto, M. Mannareli, G. Nardulli and M. Ruggieri,
Phys. Lett. {\bf B 605}, 362 (2005); \emph{ibid} {\bf B 615}, 297
(2005).

\bibitem{fuku}
K. Fukushima, Phys. Rev. D {\bf 72}, 074002 (2005).

\bibitem{alfjp05}
M. Alford and Q. H. Wang, J. Phys. G {\bf 31}, 719 (2005).

\bibitem{sch}
P. F. Bedaque and T. Sch\"{a}fer, Nucl. Phys. {\bf A 697}, 802
(2002).

\bibitem{fpi}
D. T. Son and M. A. Stephanov, Phys. Rev. D {\bf 61}, 074012 (2000);
\emph{ibid} {\bf 62}, 059902 (2000).

\bibitem{kr}
D. B. Kaplan and S. Reddy, Phys. Rev. D {\bf 65}, 054042 (2002); S.
Reddy, M. Sadzikowski and M. Tachibana, Phys. Rev. D {\bf 68},
053010 (2003).

\bibitem{kry1}
A. Kryjevski and T. Sch\"{a}fer, Phys. Lett. {\bf B 606}, 52 (2005).  In eq. 
(23) of this paper, the quasi-particles with masses $\Delta \pm 0.75 \mu_S$ have 
their minimum at $p_F - 0.25 \mu_S$ and those with masses $\Delta \pm 0.25 
\mu_S$ have their minimum at $p_F + 0.25 \mu_S$, opposite to what was written.  
We thank the authors for clarification of this typographical error.
\bibitem{kry2} A. Kryjevski and D. Yamada, Phys. Rev. D {\bf 71}, 014011
(2005).

\bibitem{forbes} M. M. Forbes, Phys. Rev. D {\bf 72}, 094032 (2005).

\bibitem{buba} 
M. Buballa, Phys. Lett. {\bf B 609}, 57 (2005).

\bibitem{sch06}
T. Sch\"{a}fer, Phys. Rev. Lett {\bf 96}, 012305 (2006).

\bibitem{QCDgaps}
D. T. Son, Phys. Rev. D {\bf 59}, 094019 (1999);
W. E. Brown, J. T. Liu and H. C. Ren, Phys. Rev. D {\bf 61}, 114012 (2000);
Q. Wang and D. H. Rischke, Phys. Rev. D {\bf 65}, 054005 (2002);
T. Sch\"{a}fer, Nucl. Phys. {\bf A 728}, 251 (2003).

\bibitem{alfj02}
 M. Alford and K. Rajagopal, J. High Energy Phys. {\bf 0206}, 031 (2002).

\bibitem{rw01}
 K. Rajagopal and F. Wilczek, Phys. Rev. Lett {\bf 86}, 3492 (2001).
\bibitem{kry03} A. Kryjevski, Phys. Rev. D {\bf 68}, 074008 (2003).

\bibitem{gorb}
E. V. Gorbar, M. Hashimoto and V. A. Miransky, Phys. Lett. {\bf B
632}, 305 (2006).

\bibitem{ins}
T. Sch\"{a}fer, Phys. Rev. D {\bf 65}, 094033 (2002).

\bibitem{alf05b}
M. Alford, P. Jotwani, C. Kouvaris, J. Kundu and K. Rajagopal, Phys. Rev. D {\bf 
71}, 114011 (2005).



\end{thebibliography}
\end{document}